\documentclass[pre,
preprintnumbers,onecolumn,11 pt]{revtex4}

\usepackage{amssymb}
\usepackage{color}
\usepackage{graphicx}

\usepackage{subfig}
\def\be{\begin{equation}}
\def\ee{\end{equation}}
\def\bea{\begin{eqnarray}}
\def\eea{\end{eqnarray}}
\usepackage{epstopdf}
\usepackage{epsfig}
\usepackage{multirow}

\begin{document}

\title{Network architectural conditions for prominent and robust stochastic oscillations}
\author{Jaewook Joo and Jinmyung Choi}
\affiliation{Department of Physics and Astronomy, University of Tennessee, Knoxville TN, 37996, U.S.A.}

\begin{abstract}

Understanding relationship between noisy dynamics and biological network architecture is a fundamentally important question, particularly in order to elucidate how cells encode and process information. We analytically and numerically investigate general network architectural conditions that are necessary to generate stochastic amplified and coherent oscillations. We enumerate all possible topologies of coupled negative feedbacks in the underlying biochemical networks with three components, negative feedback loops, and mass action kinetics. Using the linear noise approximation to analytically obtain the time-dependent solution of the master equation and derive the algebraic expression of power spectra, we find that (a) all networks with coupled negative feedbacks are capable of generating stochastic amplified and coherent oscillations; (b) networks with a single negative feedback are better stochastic amplified and coherent oscillators than those with multiple coupled negative feedbacks; (c) multiple timescale difference among the kinetic rate constants is required for stochastic amplified and coherent oscillations. 

\end{abstract}

\maketitle

\section{Introduction}
Many biological processes such as gene regulation, cellular signaling events, and metabolic reactions typically involve a large number of heterogeneous biochemical constituents and their random interactions at and across multiple temporal and physical scales~\cite{complex-g1,complex-g2}.  They can be mapped onto complex networks where a pair of constituents are linked if they interact with each other physically or chemically~\cite{complex-g3, complex-g4}.  The topological properties of the large-scale biological networks have been extensively investigated, yet, the dynamic behaviors of and on such large-scale networks remains unexplored and of great challenge despite its importance in understanding many biological timing mechanisms such as cell cycle, circadian rhythms, neural rhythms, cardiac rhythms, and hormone rhythms~\cite{cellcycle,circadian}. The technical challenges are primarily due to not only a lack of information of accurate network structure and kinetics, but also a sheer number of biochemical components and their random interactions that requires a new mathematics that can handle high-dimensional stochastic nonlinear dynamical systems. Our paper is to provide an insight into such an important stochastic nonlinear dynamical behaviors of small-size biological networks, namely coupled network motifs. 
 
Network motifs are recurring subgraphs whose appearance in certain biological networks is much more frequent than would be expected in random networks and can be considered as simple building blocks from which the network is built~\cite{motifs-g1,motifs-g2,motifs-g3}. There has been extensive research on identification of network motifs and their dynamical and functional roles in biological networks ~\cite{motifs-g2,motifs-g3,motifs-i1,motifs-i2,motifs-f1,motifs-f2}. Most of the previous studies have been limited to investigation of the dynamics of ÒbasicÓ network motifs in isolation. In real biological networks, those individual network motifs are rather coupled with each other and embedded in a larger network. Therefore, it is crucial and even imperative to study the dynamical behaviors of coupled network motifs.  
     
In Ref.~\cite{motifs-c1, motifs-c2}, Tyson {\it et. al.} classified the biochemical oscillators according to the topology of coupled negative and positive feedback loops in the underlying regulatory networks with three-components and discussed the general network architectural requirements for oscillations: negative feedback and time-delay by a series of  intermediate components. In this regard, a network with two-component negative feedback loop without explicit time-delay cannot generate oscillation, but a two-component negative feedback loop coupled with a positive feedback loop is able to generate oscillation. Here positive feedback plays a role of time-delay. Tsai {\it et. al.} uncovered an additional role of a three-component network with interlinked positive and negative feedback loops: frequency tunable oscillator~\cite{motifs-c3}. Also, it was shown that the dynamical role of coupled negative feedbacks is to realize enhanced homeostasis whereas coupled positive and negative feedbacks properly modulate signal responses and effectively deal with noise~\cite{cho1, cho2, cho3, cho4}. The previous dynamical studies on coupled feedbacks neglected stochastic fluctuations that are prevalent and sometimes dominant in intracellular biochemical processes~\cite{noise-g1,noise-g2}.  

Stochastic fluctuations can enhance the net order and induce stochastic resonance~\cite{SR-g1} where white noise amplifies a weak periodic external signal and the signal to noise ratio is maximized at the optimal noise strength. The dynamical systems that are not subject to any periodic external forcing can exhibit the similar phenomena coined as ``coherence resonance" or ``noise induced oscillations"~\cite{CR-g1,CR-g2,CR-g3,CR-g4,Mckane-1,Mckane-2, JJoo}. In the latter case, intrinsic noise can induce stochastic amplified and coherent oscillations out of a stable dynamical system. Many excitable systems are known to exhibit noise-induced oscillations~\cite{excitability, CR-g3,FN-2,HH-1} and this excitability is mistakenly assumed to be necessary for noise-induced oscillations. However, even in non-excitable systems such as a predator-prey model, particularly where the deterministic dynamical systems are stable in the entire parameter space, noise give rise to stochastic oscillations with high amplification and coherence~\cite{CR-g4,Mckane-1,Mckane-2}. Presently, there exist no theoretical work about the rigorous conditions for noise-induced oscillations, especially when the deterministic dynamical systems have no center and thus no limit cycle in the entire parameter space. 

In this paper, we analytically and numerically investigate general network architectural conditions that are necessary to generate stochastic amplified and coherent oscillations. We classify stochastic oscillators according to the topology of coupled negative feedback loops in the underlying biochemical networks with three components and furthermore identify the best network designs that can generate stochastic amplified and coherent oscillation. In the following sections, we introduce an exhaustive list of nine networks with three components, negative feedback loops, and mass action kinetics. We use the linear noise approximation to analytically obtain the time-dependent solution of the master equations and derive the algebraic expression of power spectra. The signal to noise ratio measured from a power spectrum is used to determine if the particular topology of interconnected negative feedbacks is a good design for stochastic oscillation. We find that (1) all nine networks with three-component negative feedbacks and with mass action kinetics are capable of generating stochastic amplified and coherent oscillations; (2) networks with a single negative feedback are better stochastic amplified and coherent oscillators than those with multiple coupled negative feedbacks; (3) multiple timescale difference among the kinetic rate constants is required for stochastic amplified and coherent oscillations. 

\section{Methods}

\subsection{Models}
We consider simple and idealized cell signaling networks consisting of 3 nodes ($X, Y, Z$). The nodes are the chemical species participating in the networks. The chemical reaction occurring among the nodes is represented by an edge connecting them. The reaction can be either an activation (e.g., $X \to Y$) or an inhibition (e.g., $X \dashv Y$).   

\begin{figure}[h!]
\begin{center}
\includegraphics[angle=0, width=0.75 \textheight]{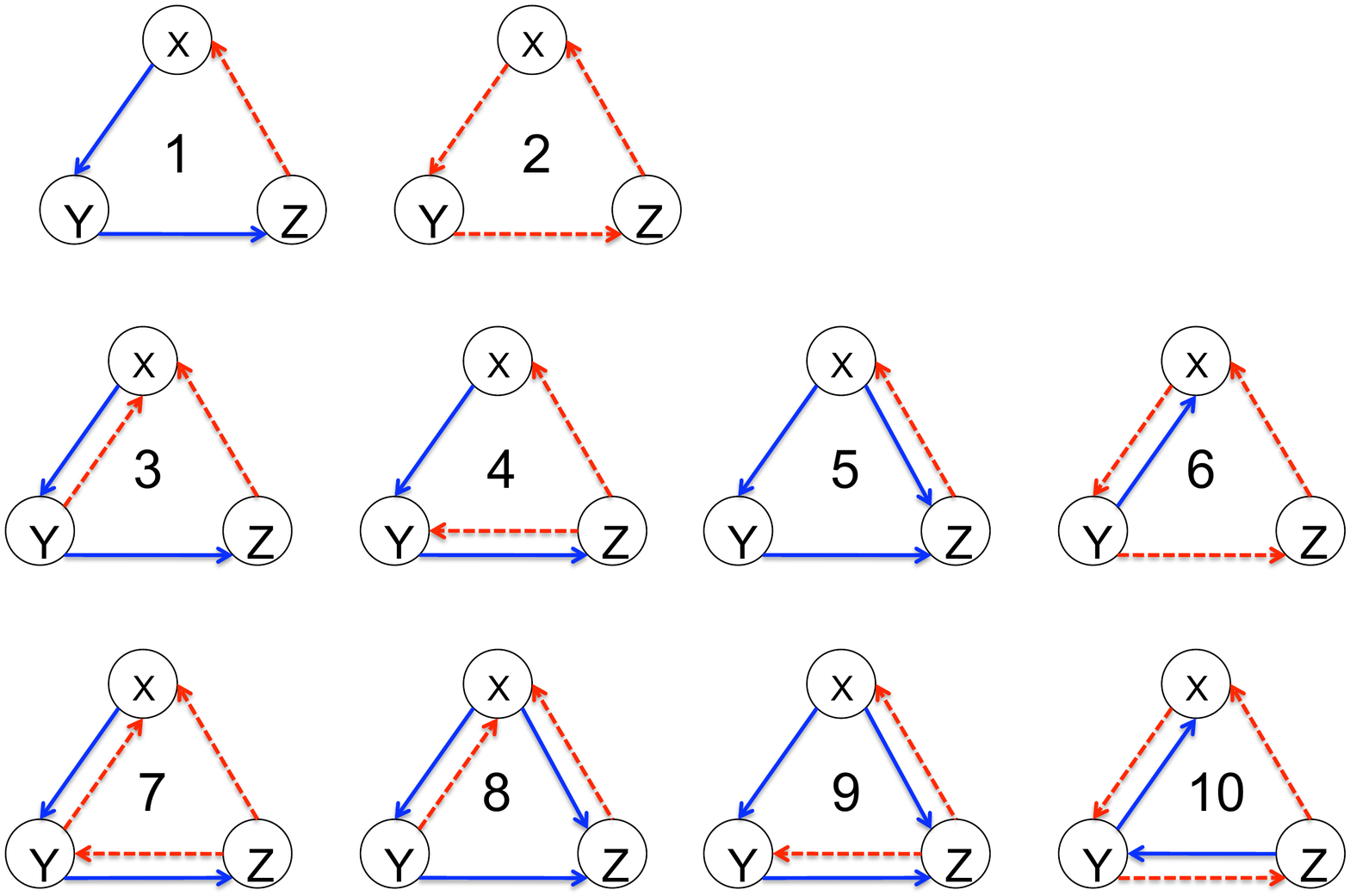}
\end{center}
\caption{Exhaustive list of all distinct networks with 3 components and negative feedback loops. The networks are numerically labeled and ordered by the number of negative feedback loops: single NFBL in the top row, two NFBLs in the middle row, and three NFBLs in the bottom row.} 
\label{Graph-nfbl}
\end{figure}

In this work, we pay a particular attention to the networks that are entirely composed of the negative feedback loops (NFBLs). The simplest networks among all are the ones with 3 nodes and 3 edges. There are only two possible configurations, i.e., i) NFBL network 1: $X \to Y \to Z \dashv X$ or ii) NFBL network 2: $X\dashv Y \dashv Z\dashv X$ so that they completes two different types of 3 component single NFBLs. If we consider more edges, we can construct multiple feedback loop networks. All the multiple NFBL networks can be built either based on NFBL network 1 or 2.  We show the names and network configurations of all the networks in Table~\ref{models} and its graphical representation in Fig.~\ref{Graph-nfbl}. They constitute the exhaustive list of the distinctive networks consisting of the negative feedback loops alone with 3 nodes. Among them, 6 multiple NFBL networks (NFBL networks 3, 4, 5, 7, 8, and 9) are based on NFBL network 1. In particular, NFBL networks 3, 4, and 5 consist of double NFBLs and NFBL networks 7, 8, and 9 triple NFBLs. There are 2 multiple loop networks (NFBL networks 6 and 10) based on NFBL network 2. NFBL networks 6 and 10 are made of double and triple NFBLs respectively.

\begin{table}[htdp]
\caption{Network Names and their topology}
\begin{center}
\begin{tabular}{|c|c|}
\hline
Names of network  &  network topology  \\
\hline
NFBL1 & $X \to Y \to Z \dashv X$   \\
\hline
NFBL2 & $X\dashv Y \dashv Z\dashv X$  \\
\hline
NFBL3 & $X \to Y \to Z \dashv X$ and $Y \dashv X$ \\
\hline
NFBL4 & $X \to Y \to Z \dashv X$  and $Z \dashv Y$ \\
\hline
NFBL5 & $X \to Y \to Z \dashv X$ and $ X \to Z$ \\
\hline
NFBL6 & $X\dashv Y \dashv Z\dashv X$  and $Y \to X$ \\
\hline
NFBL7 & $X \to Y \to Z \dashv X$, $Y \dashv X$, and $Z \dashv Y$ \\
\hline
NFBL8 & $X \to Y \to Z \dashv X$, $Y \dashv X$, and $X \to Z$ \\
\hline
NFBL9 & $X \to Y \to Z \dashv X$, $Z \dashv Y$, and $X \to Z$ \\
\hline
NFBL10 & $X \dashv Y \dashv Z \dashv X$, $X \to Y$, and $Y \to Z$ \\
\hline
\end{tabular}
\end{center}
\label{models}
\end{table}%

\subsubsection{Master Equations}
To simplify mathematical analysis, we assume that (a) any given node is constitutively synthesized. (b) any given node is degraded depending on its current abundance (c) any inhibitory interaction is described by mass action kinetics, and (d) the corresponding stochastic chemical reaction is a Markovian process. Based on these assumptions, we can write down the chemical master equations. To this end, we denote the number of each constituent of the networks $X, Y,$ and $Z$. Thus, we can describe the state of a system as a state vector, i.e., $\tilde{X} = (X,Y,Z) $. We also denote the joint probability distribution of a network in a state $\tilde{X}$ at a particular time $t$ as $P(\tilde{X};t)$ and the transition rates between two different states as $T(\tilde{X} | \tilde{X'}).$  Therefore, we can write down the master equation governing the time evolution of $P(\tilde{X};t)$ formally as 
\be
\frac{ \partial P(\tilde{X};t)}{ \partial t} = \sum_{\tilde{X'} } T(\tilde{X} | \tilde{X'}) P(\tilde{X'};t) -  \sum_{\tilde{X} } T(\tilde{X'} | \tilde{X}) P(\tilde{X};t).
\ee

The possible transitions among the states are dependent on the specific reactions allowed in each network. In Table~\ref{T-rates}, we show how to write down the transition rates of synthesis and degradation associated with all the possible chemical reactions that can occur on a single node.   

\begin{table}[htdp]
\caption{The transition rates of synthesis and degradation associated with each chemical reaction}
\begin{center}
\begin{tabular}{|c|c|c|}
\hline
Reactions  & Transition rates for synthesis  & Transition rates for degradation \\
\hline
$X$  & $k_1 \Omega$ & $k_2 X $   \\
\hline
$Y \to X$ & $k_1 Y$ & $k_2 X$  \\
\hline
$Y \to X$,  $Z \dashv X$ & $k_1 \Omega$ & $k_2 XZ/ \Omega$ \\
\hline
$Y \dashv X$,  $Z \dashv X$ & $ k_1 \Omega$ & $k_2 (XY + XZ)/ \Omega$ \\
\hline
$Y \to X$,  $Z \to X$ & $k_1(Y+Z) $ & $k_2 X$ \\
\hline
\end{tabular}
\end{center}
\label{T-rates}
\end{table}%

Based on Table~\ref{T-rates}, we can write down all the transition rules for NFBL1 as follows.   
\bea
\label{NFBL1T}
T(X +1, Y, Z | X, Y, Z) &=& k_1 \Omega. \\ 
T(X -1, Y, Z | X, Y, Z) &=& k_2 X Z/ \Omega.\nonumber \\
T(X , Y+1, Z | X, Y, Z) &=& k_3 X.  \nonumber \\
T(X ,Y-1, Z | X, Y, Z) &=& k_4 Y. \nonumber \\
T(X , Y, Z+1 | X, Y, Z) &=& k_5 Y. \nonumber \\
T(X ,Y, Z-1 | X, Y, Z) &=& k_6 Z . \nonumber
\eea

Since the number of each chemical species can be raised or lowered only by 1 in each reaction process, it is convient to introduce the step operator $E_X^{+}$ and $E_{X}^{-}$, the operation of which is defined by
\be
E_X^{\pm} f (X, Y, Z)  = f (X \pm 1, Y, Z)  = \sum_{m = 0}^{\infty} (\pm1)^{m} \frac{\partial^m}{\partial {X}^m}  f (X, Y, Z)
\ee
where $f(X,Y,Z)$ is an analytic function. Then, we can rewrite the master equation by using the step operators, 
\be
\frac{ dP(\tilde{X})}{dt} = \sum_{X_i=X,Y,Z} [E_{X_i}^{-}-1] T(\tilde{X'}|\tilde{X}) P(\tilde{X})   + \sum_{X_i=X,Y,Z}  [E_{X_i}^{+}-1] T(\tilde{X}|\tilde{X'}) P(\tilde{X'}) 
\label{M-eq}
\ee
where the sum is carried over all possible reactions raising or lowering the number of each species in the networks. 

\subsubsection{Macroscopic rate equation and Fokker-Plank equation}
The master equation involves non-linearity and for such a case, the analytic solution is not available in general. Thus, we use the system size expansion method by Van Kampen~\cite{Van-Kampen}, which allows us to expand the master equation systematically  in terms of the system size $\Omega$. Thereby, we can obtain the approximate solutions to the leading orders of the system size expansion. 
Following the standard procedure discussed in Ref.~\cite{Van-Kampen}, we define    
\be
X_i = \Omega \phi_{x_i} + \Omega^{1/2} {\xi}_i 
\label{Xtox}
\ee
where the index $i$ runs over all the chemical species, i.e., $X_i = {X,Y,Z}$. $\Omega \phi_{x_i}$ describes the macroscopic mean value of $X_i$ and the $\Omega^{1/2} \xi_{i}$ is the gaussian fluctuations of $X_i$ around its mean. Then, we can rewrite the joint probability distribution $P(\{X_i\}; t)$ in terms of the stochastic variables of $\xi_i$ such that $P(\{X_i\}) = \Pi(\{\xi_i\}) \equiv \Pi(\tilde{\xi_i}).$ According to the change of variables, the time derivative of the joint probability distribution transforms as
\be
\frac{\partial P(\tilde{X_i})}{\partial t} =  \frac{\partial \Pi(\tilde{\xi_i})}{\partial t} - \sum_{\xi_i} \Omega^{1/2} \frac{d\phi_{x_i}}{dt} \frac{\partial \Pi(\tilde{\xi})}{\partial \xi_i}. 
\ee
Consequently, the step operators are also rewritten in terms of the new variables.
\be
E_X^{\pm} \to E_{\xi}^{\pm} =\sum_{m=0} (\pm 1)^{m} \Omega^{-\frac{m}{2}} \frac{\partial^m}{\partial {\xi}^m}.  
\label{Step-OP}
\ee  
Plugging Eq.~(\ref{Xtox}) through Eq.~(\ref{Step-OP}) to Eq.~(\ref{M-eq}), we can expand the master equation systematically in the order of the system size $\Omega$. To the leading order of $\Omega$ expansion $(\Omega^{1/2})$, we obtain the macroscopic deterministic equations for all the networks. For example, we find that for NFBL1, they are given by 
\bea
\frac{d\phi_x}{dt} &=& k_1  - k_2 \phi_x \phi_y , \\ 
\frac{d\phi_y}{dt}&=& k_3 \phi_x - k_4 \phi_y ,   \\ 
\frac{d\phi_z}{dt}&=& k_5 \phi_y -  k_6 \phi_z.
\eea
We can simply repeat the same procedure for the rest of the networks to derive the macroscopic deterministic equations.  In the next leading order of $\Omega$ expansion $(\Omega^0)$, we obtain the linear Fokker-Plank Equation.  
\be
\frac{\partial \Pi}{\partial t} = - \sum_{\xi_i} \frac{ \partial}{\partial \xi_i  }C(\xi_i) \Pi + \sum_{\xi_i,\xi_j} \frac{1}{2} B_{ij} \frac{\partial^2 \Pi}{\partial \xi_i \partial \xi_j }
\ee
where $C(\xi_i) \equiv \sum_{j} J_{ij} \xi_j$. $J_{ij}$ denotes the Jacobian matrix of the macroscopic rate equations and  $B_{ij}$ is the noise covariance matrix. For NFBL1, the stability matrix and the covariance matrix evaluated at the fixed points are given by 
 \begin{equation} \label{NFBL1J}
 J =
 \left  ( \begin{array}{lll}
- k_2\phi_3^{\star} & ~~~0 & ~-k_2\phi_1^{\star}\\
 k_3 & -k_4 &~~~~0 \\
~~~0 & ~~~k_5 & ~~~-k_6  
\end{array} \right )
  \end{equation}
and 
 \begin{equation} \label{NFBL1B}
 B =
 \left  ( \begin{array}{lll}
k_1 + k_2\phi_x^{\star}\phi_z^{\star} & ~~~~0 & ~~~~0\\
~~~~0 & k_3\phi_x^{\star} + k_4\phi_y^{\star} &~~~~0 \\
~~~~0 & ~~~~0 & k_5\phi_y^{\star} + k_6\phi_z^{\star} 
\end{array} \right )
 \end{equation}
where $\phi_x^{\star}, \phi_y^{\star}$, and $\phi_z^{\star}$ refer to the fixed points of the macroscopic rate equations.

\subsubsection{Power spectrum}
In order to examine the existence of the stochastic ocillations, it is useful to transform from the time domain to the frequency domain, i.e., the Fourier transform. For this purpose, we recast the linear Fokker-Plank equation into the mathematically equivalent form of Langevin equations,
\be
\frac{d \xi_i(t)}{dt} = \sum_{j} J_{ij} \xi_j(t) + \eta_i(t)
\label{langevin}
\ee
where $<\eta_i(t) \eta_j(t')> = B_{ij} \delta(t-t').$
Taking the Fourier transform of Eq.~(\ref{langevin}),  
$
-i \omega \tilde{\xi}_i(\omega) = \sum_{j} J_{ij} \tilde{\xi}_j(t) + \tilde{\eta}_i(t)
$
where we introduce $\tilde{\xi}_i(\omega) = \int dt e^{-i \omega t} \xi_i(t)$ and $\tilde{\eta}_i(\omega) = \int dt e^{-i \omega t} \eta_i(t)$.
We then solve for $\tilde{x}_i(\omega)$ to get 
$
\tilde{\xi_i} (\omega) = \sum_{j} \Phi_{ij}^{-1}(\omega) \tilde{\eta_j} (\omega) 
$
where we define $\Phi_{ij} \equiv -i\omega \delta_{ij} - J_{ij}.$ 
Finally, we obtain the power spectrum  
 \be
 P_i(\omega) = < |\xi_i(\omega)|^2> = \sum_{j,k} \Phi_{ij}^{-1}(\omega) B_{jk} (\Phi_{ki}^{-1})^{\dagger}(\omega) = \sum_{j} \Phi_{ij}^{-1}(\omega) B_{jj} (\Phi_{ji}^{-1})^{\dagger}(\omega)
  \ee
where $\dagger$ denotes the conjugate transpose. The last equality comes from $B_{ij} $ being diagonal. Since  $(\Phi_{ji}^{-1})^{\dagger}(\omega) = (\Phi_{ij}^{-1})(-\omega)$, it further simplifies to
\be
P_i(\omega) = \sum_{j} B_{jj} (\Phi_{ij}^{-1})(-\omega) \Phi_{ij}^{-1}(\omega) B_{jj} =  \sum_{j} B_{jj} |\Phi_{ij}^{-1} (\omega)|^2. 
\ee 
For 2 by 2 $J$ and $B$ matrices, the algebraic expression of the power spectrum is simple and according to Ref.~\cite{Mckane-1}, it is given by  
\be
P_i(\omega) = \frac{a_i \omega^2 + b_i}{ \omega^4 + \alpha\omega^2 + \beta  }
\ee
where $i=1, 2$, and 
\bea
\label{psd-coef}
a_1 =B_{11}, &~~& b_1 = B_{11} J_{22}^2 + B_{22} J_{12}^2, \\
a_2 =B_{22}, &~~& b_2 = B_{11} J_{21}^2 + B_{22} J_{11}^2, \nonumber 
\eea
and $\alpha = Tr[J]^2 - 2Det[J]  , ~~ \beta= Det[J]^2.$

In particular, the condition for the existence of power spectrum peak is equivalent to that of $z > 0$ such that  $dP_i(z)/dz =0 $ where $z = \omega^2$: 
$a_iz^2 + 2b_iz + \alpha b_i - a_i \beta =0$. 
According to the Descartes' rule~\cite{Murray}, the condition for the quadratic equation to have a real positive root is given by 
\be
\alpha b_i - a_i \beta < 0~~~\Leftrightarrow ~~~~(Tr[J]^2 - 2Det[J]) b - a Det[J]^2  < 0
\label{peak-cond-2d}
\ee
where $a_i > 0$ and $b_i > 0$ are trivially satisfied as shown in Eq.~(\ref{psd-coef}).

For our networks, $J$ and $B$ are given by 3 by 3 matrices, the explicit expression of the power spectrum for $i=x,y,z$ in terms of matrix elements of $J$ and $B$ is given by 
\be 
P_i(\omega) =\frac{ \beta_1\omega^4 + \beta_2 \omega^2 + \beta_3 }{\omega^6 + \alpha_1\omega^4 + \alpha_2\omega^2 + \alpha_3}
\label{Pw}
\ee  
where the coefficients  in the numerator are: $\beta_1= B_{ii}$, $\beta_2 = B_{ii}[ (Tr[J]-J_{ii})^2 -2M_{ii} ] + B_{i+1,i+1}J_{i,i+1}^2 +B_{i+2,i+2}J_{i,i+2}^2$, and $\beta_3= B_{ii}M_{ii}^2 + B_{i+1,i+1}M_{i,i+1}^2+ B_{i+2,i+2}M_{i,i+2}^2$ and $M_{ij}$ denotes the minor matrix of $J_{ij}$. The index of the matrices obeys the following conventions: $J_{i,j}=J_{i+3,j+3}$, $B_{i,j}=B_{i+3,j+3}$, and $M_{i,j} = M_{i+3,j+3}$. The coefficients in the denominator are: $\alpha_1 = Tr[J]^2-2(J_{11}J_{22}+J_{22}J_{33}+J_{33}J_{11} - J_{12}J_{21} -J_{23}J_{32} - J_{31}J_{13}) $, $\alpha_2 = (J_{11}J_{22}+J_{22}J_{33}+J_{33}J_{11} - J_{12}J_{21} -J_{23}J_{32} - J_{31}J_{13})^2 - 2Tr[J]Det[J]$, and $\alpha_3 = Det[J]^2$.

\subsection{Linear Stability Analysis}
The set of deterministic rate equations for each network takes the general form of
\be
\label{gODEs}
\dot{\phi_{x_i}} = f_{i}(\phi_{x},\phi_{y},\phi_{z}).
\ee
For such a system, determining the stability of the steady state, $\dot{\phi_{x}} = \dot{\phi_{y}} = \dot{\phi_{z}} = 0$, entails the examination of the linearised form of ~(\ref{gODEs}).  Namely, in performing a multivariate Taylor expansion about the fixed points, $\{\phi_{x}^{\star},\phi_{y}^{\star},\phi_{z}^{\star}\}$, the above system of non-linear ODEs may be approximated by the first order term which is linear with respect to $\delta \phi_{\imath} = (\phi_{\imath} - \phi_{\imath}^{\star}$).  More explicitly,
\begin{equation}
\left(\begin{array}{c}
\dot{\delta \phi_{x}}\\
\dot{\delta \phi_{y}}\\
\dot{\delta \phi_{z}}
\end{array}\right) = \left(\begin{array}{ccc}
J_{11} & J_{12} & J_{13}\\
J_{21} & J_{22} & J_{23}\\
J_{31} & J_{32} & J_{33}
\end{array}\right)_{ \phi_{x}^{\star}, \phi_{y}^{\star},  \phi_{z}^{\star}} \left(\begin{array}{c}
\delta \phi_{x}\\
\delta \phi_{y}\\
\delta \phi_{z}
\end{array}\right)
\end{equation}
where $J_{\imath \jmath} = \frac{\partial f_{\imath}}{\partial \delta \phi_{\jmath}}\arrowvert_{{\phi_{x}^{\star},\phi_{y}^{\star},\phi_{z}^{\star}}}$ represents the $\imath^{th},\jmath^{th}$ element of the Jacobian matrix.  According to the Hartman-Grobman theorem, so long as the eigenvalues of the Jacobian do not lie on the imaginary axis, the linearised form may be used to assess the stability of the steady state for the non-linear system.  In light of this fact, we begin then by grouping each feedback loop into a family derived from either NFBL1 or NFBL2.  In doing so, one notices for NFBL2, NFBL6, and NFBL10, the Jacobian takes on the general form
\begin{equation}
J = \left(\begin{array}{ccc}
-a_{11} & a_{12} & -a_{13}\\
-a_{21} & -a_{22} & a_{23}\\
0 & -a_{32} & -a_{33}
\end{array}\right) 
\end{equation}           
for $a_{\imath \jmath} = \arrowvert J_{\imath\jmath} \arrowvert$ being greater than zero so long as each fixed point is greater than zero. 
A proof of stability for the deterministic system amounts to demonstrating $Re(\lambda)<0$, where $\lambda$ denotes the eigenvalue of the Jacobian matrix.  Finding such eigenvalues requires one to solve for the characteristic equation of the $ 3 \times 3$ determinant, $ det(J - I \lambda) = 0$, given here by
\begin{equation}
\lambda^{3}-Tr(J)\lambda^2+(A_{11}+A_{22}+A_{33})\lambda - det(J) = 0
\end{equation}
for $Tr(J)$ representing the trace of the Jacobian, while $A_{\imath \jmath}$ is the cofactor of the $\imath^{th}$ row and $\jmath^{th}$ column.  By applying the Routh-Hurwitz conditions for a third order characteristic equation, it is guaranteed $Re(\lambda)<0$ so long as  $Tr(J)<0, ~~ det(J)<0, ~~det(J)-(A_{11}+A_{22}+A_{33})Tr(J)>0$.
\bea
&&det(J)-(A_{11}+A_{22}+A_{33})Tr(J)  \\
&=& a_{11}a_{12}a_{21}+a_{12}a_{21}a_{22}+a_{22}a_{23}a_{32}+a_{23}a_{32}a_{33} \nonumber \\ 
&+&a_{11}^{2}a_{22}+a_{11}a_{22}^{2}+a_{11}^{2}a_{33}+a_{22}^{2}a_{33} \nonumber \\ 
&+&a_{11}a_{33}^{2}+a_{22}a_{33}^{2}+2a_{11}a_{22}a_{33}-a_{13}a_{21}a_{32}.  \nonumber
\eea
Interestingly enough, such an expression will always positive provided $2a_{11}a_{22}a_{33}-a_{13}a_{21}a_{32}>0$.  For each network derived from NFBL2, it turns out 
$2a_{11}a_{22}a_{33}-a_{13}a_{21}a_{32} = k_{2}k_{4}k_{6}\phi_{x}^{\star}\phi_{y}^{\star}\phi_{z}^{\star}$   
for any general case.  In noting this quantity to always be positive, it follows, that all networks in the family of NFBL2 are stable.  
For the family of the networks stemming from NFBL1, it can be shown that all three Ruth-Hurwitz conditions always hold. Consequently, all NFBL networks presented have stable steady state solutions.

\subsection{Numerical methods}
In the following, we explain the details of sampling methods, power spectrum and SNR calculations. We also explain how we define the robustness and the prominence of the stochastic oscillations for our study. 

\subsubsection{Sampling methods}
As shown in the stability analysis, our networks are entirely stable for positive real domain in the parameter space, i.e., $\{k_{i}\} \in R^{+}.$ In order to demonstrate the existence of the stochastic amplified oscillations, we sample $10^6$ sets of the 5 kinetic rate constants (setting $k_2=1$) uniformly and randomly from the logarithmically scaled interval [$10^{-3}$, $10^3$] by using the unconstrained Monte Carlo sampling technique. We also used three other different sampling methods: MC sampling from linear parameter space, LHS (Latin Hypercube sampling) from linear and logarithmic parameter space. Both MC and LHS Sampling from linear parameter space didn't provide a good number of SNR$>1$. The LHS sampling from logarithmic parameter space produced the very similar results as the MC sampling from logarithmic parameter space.

\subsubsection{Power Spectrum and SNR calculation}
The SNR (signal to noise ratio) is commonly defined by the power spectrum peak height over its relative width, i.e., 
\be
SNR = \frac{P(\omega_o)}{\Delta \omega/ \omega_o}
\ee
where $\omega_o$ and $P(\omega_o)$ denote the peak frequency and the peak height respectively. The $\Delta \omega$ defines the so-called full width at half maximum (FWHM).  
Our calculation of the power spectrum and the SNR is based on its analytic expression obtained from the earlier section. It is, however, not straightforward to obtain the analytic expression for SNR because SNR calculation requires the solution of 4th order polynomial equation given by $dP(z)/dz = 0$ where $z = \omega^2$ and the determination of $\Delta \omega$ around $\omega_o$ is also non trivial. Therefore, we use a numerical algorithm to calculate the SNR in the following way. 
First, we identify the occurrence of the maxima of $P(\omega)$ for a given set of the kinetic rate constants and fixed points. In this step, we obtain the numerical values of $P(\omega)$ from the analytic expression of Eq.~(\ref{Pw}) and then examine the condition for the existence of maxima. For example, $P(\omega)$ values are evaluated at between $\omega$ = 0 to 200 in the increment of $\Delta \omega =0.01$. In each increment, we calculate the difference of $\Delta P_1 = P(\omega) - P(\omega-\Delta \omega)$ and $\Delta P_2 = P(\omega + \Delta \omega) - P(\omega)$. If $\Delta P_1 > 0$ and $\Delta P_2< 0$, we locate the peak frequency at $\omega = \omega_o.$  Second, if the peak frequency $\omega_o$ exists, we search for the two neighboring frequencies, i.e., $\omega_1 (< \omega_o )$ and $\omega_2 (>\omega_o)$ that satisfy
 $P(\omega_1) = P(\omega_2) = P(\omega_o)/2$. Then, we can determine $\Delta \omega$ from the difference of two frequencies, i.e., $\Delta \omega = |\omega_2 - \omega_1|$.        

\subsubsection{Probability distribution of SNR}     
The probability distributions of SNR is obtained from the histogram of SNR values in the range of 0 to $10^7$ followed by the normalization. To obtain the histogram, we set the bin size to vary in the linear scale to smooth out the noisy curves, which occur due to the fluctuations in the number of data sets as SNR values get larger. For example, in the range between SNR =1 and 10, the bin size is set to 1 and in the range between SNR =10 and 100, the bin size = 10.

\subsubsection{Robustness vs. Prominence}
In this work, the robustness of a networks is defined as the size of the domain of the parameter space that generates the stochastic amplified oscillator, i.e., SNR$>1$ whereas the prominence is the average value of SNR within such domain for that networks.  We measure the domain of the parameter space from counting the data sets showing SNR $>1$ out of total sampling points and calculate the average value of SNR within the domain by integrating the probability distribution of SNR, i.e.,
\be
\langle {SNR} \rangle = \sum_{i=x,y,z} \frac{1}{ D_i} \int P_i(SNR) \Theta(SNR - 1) d(SNR)
\ee
where $D_i$ denotes the magnitude of the domain that show SNR $>1$ for a given species $i=x,y,z$. So $D_i$ is equivalent to the counts of the data sets with SNR $>1$ for a given species in a model. The summation over $x,y,z$ is carried out only among those species that exhibit SNR$>1$. For example, NFBL3 do not show SNR $>1$ for $x$ species. Thus the summation is limited to $y$ and $z$ species. The step function $\Theta(x)$ is inserted to restrict the integral range only for SNR$>1$.  

\section{Results}

\subsection{All the networks with 3 nodes and mass action kinetics  and only negative feedback loops are capable of being stochastically amplified oscillators.} 

\subsubsection{ Network with two components, a negative feedback, and mass action kinetics}

\begin{figure}[h!]
\begin{center}
\includegraphics[angle=0, width=0.75 \textheight]{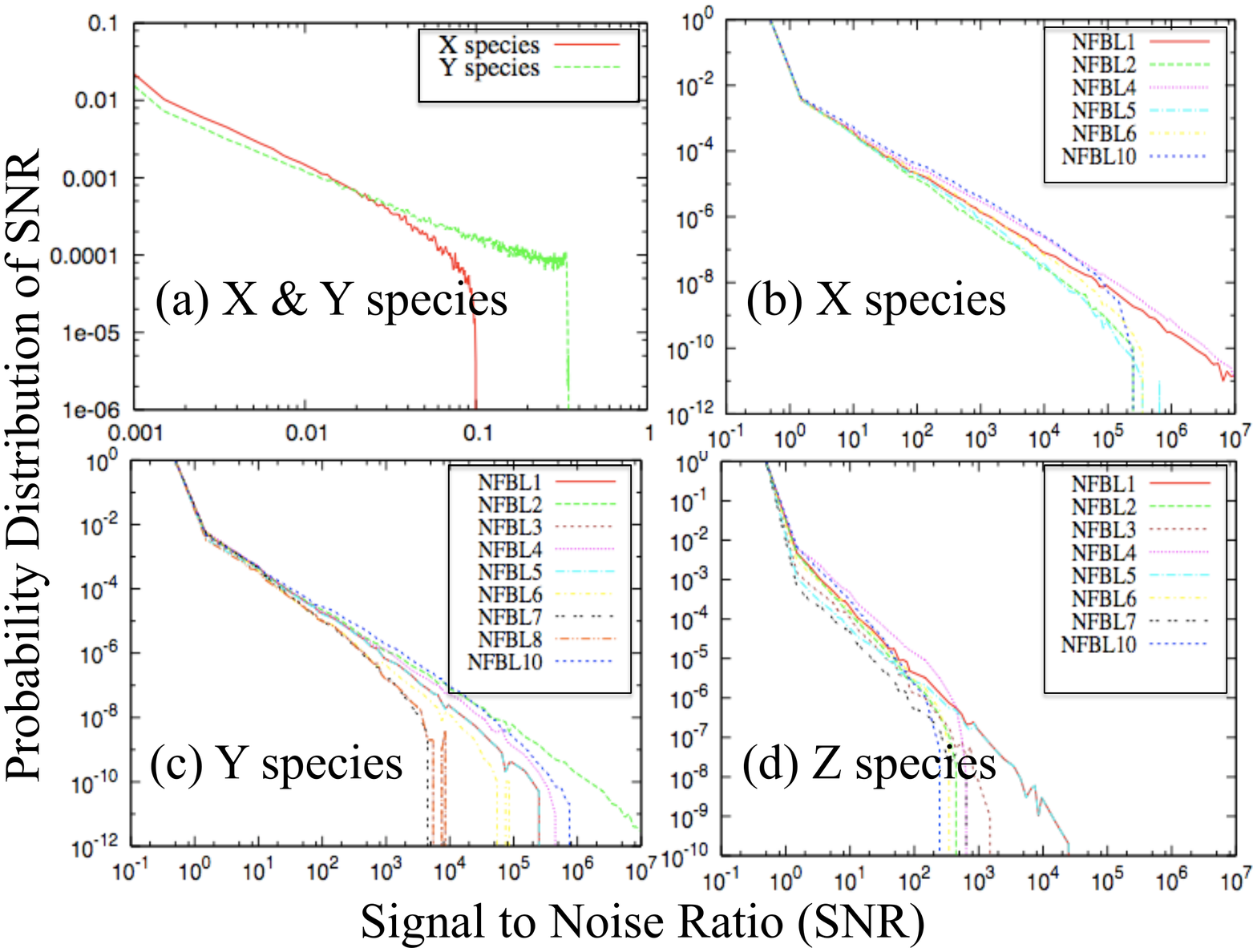}
\end{center}
\caption{Probability distribution of Signal to Noise Ratio (SNR) for (a) a network with two-component negative feedback loop and (b)-(d) nine networks with three-component negative feedback loop. The SNR distributions for both species X and Y are plotted in (a) whereas those for X, Y, and Z species are plotted in (b), (c), and (d), respectively.  We note that not all species of a network necessarily exhibit SNR$>$0.} 
\label{SNR-histo}
\end{figure}

The networks we consider in this work are built upon the following constraints: i) The networks have mass action kinetics ii) No positive feedback loop. iii) 3 nodes.  The third constraint, 3 nodes, is not a necessary condition for stochastic oscillations as discussed in Ref.~\cite{Mckane-1, Mckane-2}. But, if we add the constraints i) and ii), having 3 nodes appears to be a necessary architectural condition to build a good stochastic oscillator. We can show that NFBL with 2 nodes is not a good stochastic oscillator.

\begin{figure}[h!]
\begin{center}
\includegraphics[angle=0, width=0.75 \textheight]{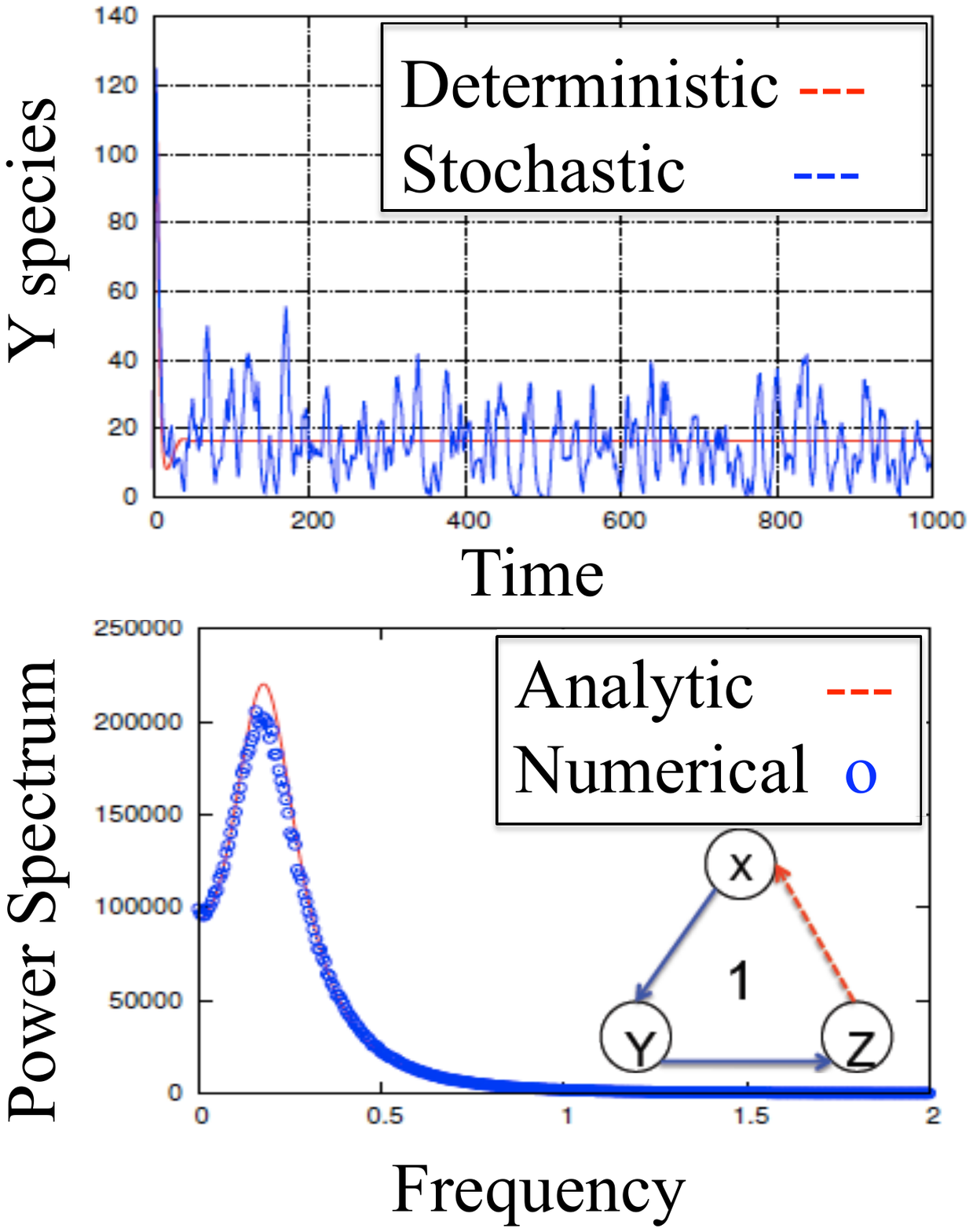}
\end{center}
\caption{Top panel: comparison between the stochastic time series and the deterministic time series for Y species of the NFBL network 1. Bottom panel: comparison between two power spectra, one from the linear noise approximation and another from the numerical simulation.} 
\label{Stochastic-sim}
\end{figure}

We consider a network with 2 component NFBL constructed with the same design rules used for the 3 component NFBL network counterpart: $X \to Y$ and $Y \dashv X$. 
For the detailed discussion of the 2 component network and the derivation of its power spectrum, we refer the readers to the methods section and Ref. [27]. We find that there exist some sets of kinetic rate constants that satisfy Eq~(\ref{peak-cond-2d}), i.e., the condition for the existence of the peak of the power spectrum. But, having a peak does not necessarily mean the existence of highly amplified and coherent stochastic oscillations. In Fig.~\ref{SNR-histo} (a), we show the probability distribution of SNR for $X$ and $Y$ species for the 2 component NFBL network. We sample $10^6$ sets of the kinetic rate constants in the logarithmic scale from $10^{-3}$ to $10^3$. The SNR values appear to be bounded below 1. This numerically proves that the 2 component NFBL network  is not a good model to obtain sufficient amplified stochastic oscillations.

\subsubsection{Networks with three components, negative feedbacks, and mass action kinetics}

In contrast to the 2 component NFBL network, the 3 component NFBL networks exhibit highly amplified and coherent stochastic oscillations. In Fig.~\ref{Stochastic-sim} (a), we compare a stochastic time series data with the corresponding deterministic trajectory obtained from solving the macroscopic rate equations for NFBL network 1. The time series data don't appear to have clearly discernable periodic behavior. However, as shown in Fig.~\ref{Stochastic-sim} (b), we can see that the power spectrum obtained from averaging over 1000 realizations of such stochastic time series data shows a prominent peak with small width, which is the evidence of the periodicity of stochastic time series data. It also verifies that the power spectrum derived from the linear noise approximation is in good agreement with that simulated with GillespieÕs direct algorithm in a carefully chosen systems size.

As shown in the Fig.~\ref{SNR-histo} (b), (c), and (d), all the networks with 3 component NFBL exhibit the stochastic oscillations quantified by the SNR in a broad range from 0 to $10^7$. All 9 networks are deterministically stable in the entire parameter space but noise is capable of inducing highly amplified and coherent stochastic oscillations. It appears that the SNR probability distributions for all the networks follow a similar trend of a power-law behavior followed by a cut-off. We have not investigated if the cut-off is dependent on the sampling size within the biologically relevant parameter value range.

\subsection{Networks with a single negative feedback loop are better stochastic oscillators than those with multiple coupled negative feedback loops} 

As shown in the previous section A, all networks are capable of being stochastic simplified and coherent oscillators if the kinetic rate constants are properly balanced. In this section, we compare those 9 networks and identify the network architectural conditions for, namely the design principles of, the most prominent and robust stochastic oscillators.

\begin{figure}[h!]
\begin{center}
\includegraphics[angle=0, width=0.75 \textheight]{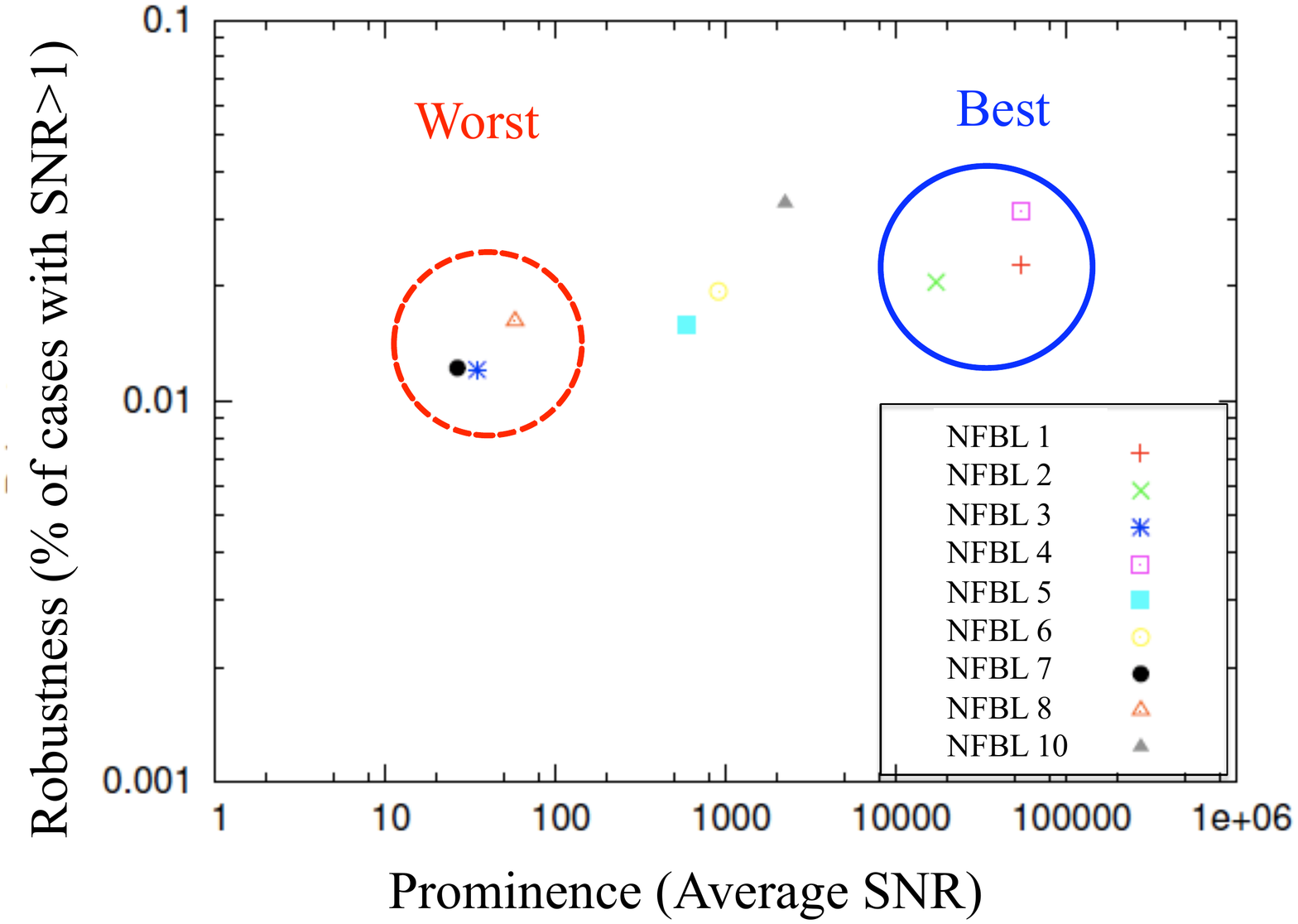}
\end{center}
\caption{Network design space for identification of stochastically prominent and robust oscillators. Robustness is plotted against prominence. Please refer to both main text and methods for the detailed definition of robustness and prominence.} 
\label{RP}
\end{figure}

In Fig.~\ref{RP} is shown the prominence versus  the robustness of all nine networks. For our particular networks, the prominence is defined as the average value of the signal to noise ratio (SNR) over the stochastic oscillators with SNR$>1$ and over three variables ($X, Y,$ and $Z$). In this prominence measure, we exclude the cases with SNR = 0 and SNR$<1$ because the prominence is supposed to measure how good the stochastic oscillators are when they satisfy the noise-induced oscillation conditions: the existence of a peak of a power spectrum and the signal being greater than stochastic fluctuations, i.e., SNR$>1$. The robustness is measured by the average percentage of the cases of SNR$>1$ over the total sampling size. Each network has three nodes ($X, Y,$ and $Z$) and all three variables are not necessarily stochastically oscillatory nor have the same average SNR. For example, a variable Y of a network can exhibit a highly amplified stochastic oscillation while a variable $X$ of that network cannot have a single case of SNR$>1$ among the total samples. In Fig.~\ref{RP}, we present the prominence and the robustness averaged over the cases of SNR$>1$ and over the three variables of a network. Please note that two measures are similar to the maximum of the average SNR among three variables. Nine networks are classified into three distinctive groups, based upon the prominence measure. The NFBL networks 3, 7, and 8 are categorized as the worst performers, the NFBL networks 1, 2, and 4 as the best performers, and the NFBL networks 5, 6, and 10 as the mediocre performers. Between the best and the worst performers, there is a large difference in the prominence measure, by 3 to 4 orders of magnitude.

We identify the common network architecture among the worst performers: the NFBL networks 3, 7, and 8 have ÒcompetingÓ double inhibitions acting on a node $X$ as shown in Fig 1. For all of the worst performers, the node X has zero case with SNR$>1$ out of the total sample size $10^{6}$. When both inhibitions act on the same node in an asynchronous and random manner, they compete to influence the node X in different directions, resulting in cancelling out of their driving forces on the node. This is similar to the case of a pendulum perturbed by two random forces, whose forces are completely cancelled out on average. The similar situation can be found in deterministic oscillators. 

All of the worst performing networks belong to the family of the networks derived from NFBL network 1 which can be generated by adding one or more directed inhibition or activation edges to NFBL network 1 with a single constraint that any additional edge should not create a positive feedback loop. Thus, it is topologically allowed in this family of NFBL network 1 to have double inhibitions or double activations acting on the same node.  However, the family of networks stemming from NFBL network 2 is in principle not allowed to have any ÒcompetingÓ double inhibitions or activations. Any additional directed edges introduced to the NFBL network 2 are allowed to be only an activation. This architectural constraint make this latter family of NFBL networks either the best or the mediocre stochastic amplified oscillators. It is notable that the same network architecture as NFBL network 2 was used to demonstrate the synthetically designed gene regulatory oscillations in noisy intracellular setting in Escherichia coli~\cite{EL}.

\subsection{Multiple timescale difference among the kinetic rate constants is required for stochastic amplified and coherent oscillations.} 
The kinetics that synthesize and degrade the interacting biochemical species must occur on appropriate timescales that permit a network with three-component negative feedback loop to generate the stochastically amplified and coherent oscillations.

\begin{figure}[h!]
\begin{center}
\includegraphics[angle=0, width=0.75 \textheight]{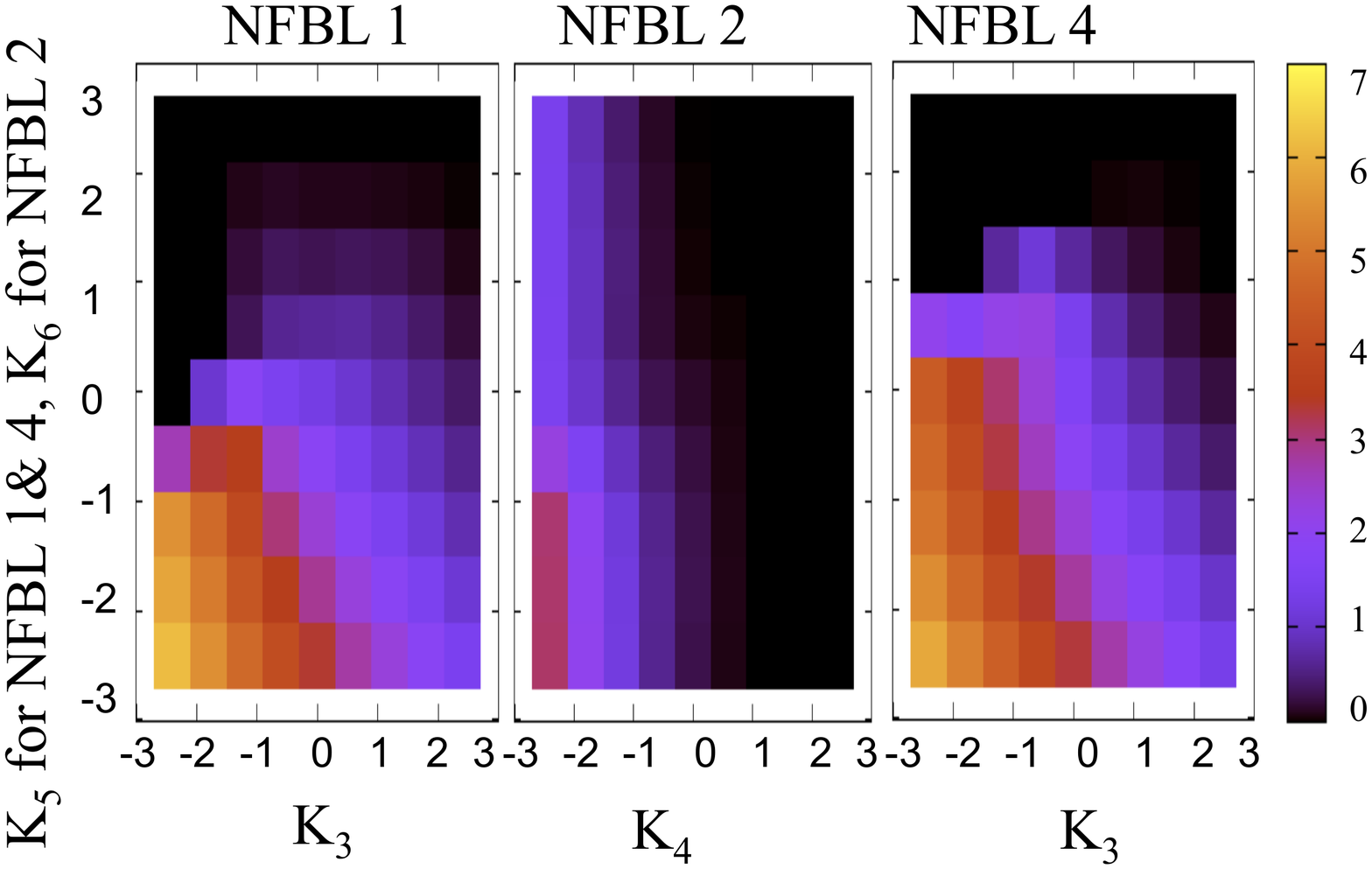}
\end{center}
\caption{Phase diagram of Signal to Noise Ratio for X species of three NFBL networks that generate the most prominent and robust stochastic oscillations. Those SNR vales are averaged over 3 kinetic rates constants and coarse-grained.} 
\label{2d-histo}
\end{figure}

In Fig.~\ref{2d-histo} are shown the two-dimensional phase diagrams for the X species of three best performing networks, NFBL networks 1, 2, and 4. The SNR values are averaged over three other kinetic rate constants and mapped onto the two-dimensional parameter space. For all three networks, the highest average SNR values are located around the left bottom corner of the diagrams where both kinetic rate constants are in the order of $10^{-2}$ to $10^{-3}$. The similar distribution of the average SNR values are found, but not presented here, for the X species of three other networks with mediocre performance, NFBL networks 5, 6, and 10. As for the Y species, all six networks exhibit the highest average SNR values around the right bottom corner where one kinetic rate constant is in the order of $10^{3}$ and another in the order of $10^{-3}$, thus indicating their ratio in the order of $10^{6}$. 

Since we rescale the time with $k_2$ (i.e., setting $k_2$=1), the other rate constants are effectively normalized by $k_2$. The characteristic distribution of the average SNR in two dimensional phase diagrams in Fig.~\ref{2d-histo} indicates that it is required for the stochastic amplified and coherent oscillations that both kinetic rate constants $k_3$ and $k_6$ for NFBL networks 1 and 4 (or $k_4$ and $k_6$ for NFBL networks 2) should be by two to three order of magnitude smaller than the kinetic rate constant $k_2$. Both $k_3$ and $k_5$ are the synthesis rate of Y and Z species and $k_2$ is the degradation rate of X species. Thus, this finding suggests that there must be a multiple timescale requirement for stochastic oscillations similar to that found in Ref.~\cite{motifs-c1}, i.e., proper balancing among rates of synthesis and degradation of interacting chemical species for a network with three-component negative-feedback loop to generate oscillations. For NFBL network 2, all three rates $k_2$, $k_4$, and $k_6$ are the degradation rates induced by inhibition. This also hints that in order for NFBL network 2 to generate the stochastic oscillations, there must be multiple timescale difference among the degradation rates.

 \section{Discussion}
 
The goal of this paper is to investigate the relationship between network structure and stochastic dynamical behaviors of small-size biological networks with coupled negative feedbacks. We identify the design principles of the best stochastic amplified and coherent oscillators that are made of only negative feedbacks: three components, a single or at most two negative feedbacks, and no competing double-inhibitions acting on a node. In other words, the more number of negative feedbacks a network has, the more likely it has double-inhibitions, and thus the less likely it exhibits stochastic oscillation. This design agrees with the common network architectures found among biological oscillators. E.g., the core part of the fungal circadian clock operates with a single transcription/translation negative feedback loop where a single gene $\it frq$ inhibits its own expression in a stochastically cyclic manner~\cite{circadian}. There is a counter example, too. Nucleo-cytoplasmic shuttling of NF-$\kappa$B is regulated by multiple coupled negative feedbacks such as I$\kappa$B$\alpha$, I$\kappa$B$\beta$, I$\kappa$B$\epsilon$, and A20. This signaling network is not the best stochastic oscillator according to our work, but exhibits noisy oscillations in single cells~\cite{nelson}. One possible explanation might be that the NF-$\kappa$B signaling network is coupled with a positive feedback through autocrine signaling of TNF$\alpha$ and thus should be regarded as the interconnected positive and negative feedbacks which is one of the most common coupled feedbacks in biological oscillators.

Since any intracellular biochemical reactions are subject to random noise, a plethora of noise-induced oscillation phenomena have been found in biological oscillations and extensively investigated for over 30 years~\cite{circadian,CR-g2,CR-g3,CR-g4,excitability,FN-2,HH-1,Mckane-1,Mckane-2,JJoo,Van-Kampen, EL, nelson}. However, there exist not much theoretical work available on the necessary and sufficient conditions for noise-induced oscillations~\cite{CR-g2, Mckane-1}. In this paper, we are intrigued by a similar, but biologically relevant question: what are the necessary network architectural conditions for noise-induced oscillations? Our numerical work uncovers one of those conditions, a single negative feedback. But, we are left with many more questions. What are the precise mechanisms that make multiple negative feedback loops detrimental to noise-induced oscillation? How can noise induce oscillation when the deterministic dynamical systems have no center and thus no limit cycle in the entire parameter space? One extreme case related to the latter question is very intriguing. Our unpublished data show that the signal to noise ratio, a metric to judge noise-induced oscillation, can be very high even though all of the eigenvalues of the stability matrix are real and negative. In this case, the deterministic system does not operate with any natural frequency, but some unknown mechanisms govern and determine a timing (or resonant frequency) of noise-induced oscillation.

In this paper, we consider a small set of networks with three components, negative feedback loops, and mass action kinetics. The above three constraints are to simplify and ease our modeling exercise and thus reflect the limitations of our models. In our future work, we relax those constraints to investigate networks with {\it N} components, both positive and negative feedback loops, and/or nonlinear kinetics. From our comparative studies between two-component network and three-component networks, we can conjecture that the larger number of intermediate components a negative feedback network has, the more time-delay it has, and the more likely it exhibits stochastic oscillation. This conjecture can be tested with a stochastic version of N-cyclic network with negative feedbacks~\cite{tyson,arcak}. Our current modeling framework can be easily modified and extended to enumerate networks with interconnected positive and negative feedbacks and identify the most robust and prominent stochastic oscillators within this group of networks.

\end{document}